\pdfoutput=1
\documentclass[pra,showpacs,lengthcheck]{revtex4-1}

\usepackage{graphicx}
\usepackage{amsmath}
\usepackage{amssymb}

\newcommand{\ket}[1]{\left|#1\right\rangle}
\newcommand{\bra}[1]{\left\langle#1\right|}
\newcommand{\ip}[2]{\left\langle#1\left|\right.#2\right\rangle}
\newcommand{\ev}[1]{\left\langle#1\right\rangle}

\newcommand{\bd}{\partial}
\newcommand{\veps}{\varepsilon}

\begin{document}

\title{Analysis of non-Markovian coupling of a lattice-trapped atom to free space}

\author{Michael Stewart}
\email{michael.stewart@stonybrook.edu}
\author{Ludwig Krinner}
\author{Arturo Pazmi\~no}
\author{Dominik Schneble}
\affiliation{Department of Physics \& Astronomy, Stony Brook University, Stony Brook, NY 11794-3800, U.S.A.}
\date{\today}

\begin{abstract}
Behavior analogous to that of spontaneous emission in photonic band gap materials has been predicted for an atom-optical system consisting of an atom confined in a well of a state-dependent optical lattice that is coupled to free space through an internal-state transition [de Vega \textit{et al.}, Phys. Rev. Lett. \textbf{101}, 260404 (2008)]. Using the Weisskopf-Wigner approach and considering a one-dimensional geometry, we analyze the properties of this system in detail, including the evolution of the lattice-trapped population, the momentum distribution of emitted matter waves, and the detailed structure of an evanescent matter-wave state below the continuum boundary. We compare and contrast our findings for the transition from Markovian to non-Markovian behaviors to those previously obtained for three dimensions.
\end{abstract}

\maketitle

\section{Introduction}
\label{SEC:intro}

The description of spontaneous emission is a fundamental topic in quantum optics.
The paradigmatic example of an excited two-state atom coupled to a continuum of empty photon modes and undergoing exponential radiative decay was first analyzed by Weisskopf and Wigner \cite{Weisskopf1930}, while Purcell later showed that the spectral density of modes available for photon emission plays a crucial role in determining its time evolution \cite{Purcell1946}. In the most extreme case, narrowing the spectrum towards a single mode allows for the attainment of the strong-coupling regime of cavity QED \cite{Meystre2007} with coherent Rabi-oscillations, but already weaker modifications of the vacuum can have dramatic and novel effects.

One example for such a modification is provided by photonic band gap (PBG) materials \cite{John1984, John1990, John1991}, where the assumption of an unbounded spectral function \cite{Lodahl2004} made in the original Weisskopf-Wigner model is broken. In PBG materials, spectral gaps can lead to strong deviations from Markovian (exponential) decay \cite{Lewenstein1988, Kofman1994, Bay1997, Lambropoulos2000, Berman2010B, Yablonovitch1987}. Moreover, for optical transitions that lie in such a gap, the coupling to the mode continuum is predicted to result in dressed, so-called atom-photon bound states \cite{John1990, John1991, Kofman1994, Lambropoulos2000, Bykov1975}, in which the emitting atom is surrounded by a localized cloud of photonic excitations. Such bound states in PBG materials (and also in waveguide QED devices \cite{Calajo2016}) are expected to be robust to dephasing \cite{Mogilevtsev2005A}, which should make them attractive for quantum computation \cite{Mogilevtsev2005B}.

As first suggested in \cite{deVega2008, Navarrete_Benlloch2011}, the main features of a radiating atom coupled to a PBG material can also be implemented and studied in an atom-optical setting, where an atom confined to a single well of a deep optical lattice is coupled to free space via an internal-state transition. Here, the occupational spin represented by the presence and absence of the trapped atom in the well corresponds to the ground and excited state of the radiating atom in the optical case (following terminology first introduced  in \cite{Recati2005}, one may call the emitting well an ``atomic quantum dot''). 
Using the Weisskopf-Wigner approach, we give an in-depth analysis of this system for the case of a one-dimensional, tube-like geometry. We focus on the vicinity of the bound-unbound transition, which we find is accompanied by a strong shift from Markovian to non-Markovian dynamics in the form of long-lived damped oscillations. In addition to the population dynamics, we characterize the momentum distribution of the emitted matter waves. We give a detailed analysis of the structure of evanescent matter waves that arise for coupling below the edge of the continuum, and the results of which depart strongly from the behavior predicted \cite{deVega2008, Navarrete_Benlloch2011} for three dimensions.  

This paper is structured as follows: section \ref{SEC:atomicmodel} develops the Weisskopf-Wigner Hamiltonian for our system. In section \ref{SEC:markovcase} we analyze the dynamics of the lattice-trapped population and the transition from Markovian to non-Markovian behavior. Results on the momentum distributions of the emitted matter waves and their dependence on the coupling parameters are presented in section \ref{SEC:momentumdistributions}. The structure of the evanescent matter wave surrounding the well below the continuum edge is the topic of section \ref{SEC:boundpairs}. The relationship between the evanescent matter wave state and the concept of an atom-photon bound state in a PBG material is explored in section \ref{SEC:analogapbs}. We conclude in section \ref{SEC:conclusion} with some experimental considerations. The appendix \ref{APP:WignerWeisskopf} connects the Markovian case with the standard results of Weisskopf-Wigner theory for spontaneous emission.

\section{Weisskopf-Wigner Hamiltonian}
\label{SEC:atomicmodel}

\begin{figure}[h!]
\centering
    \includegraphics[width=0.95\columnwidth]{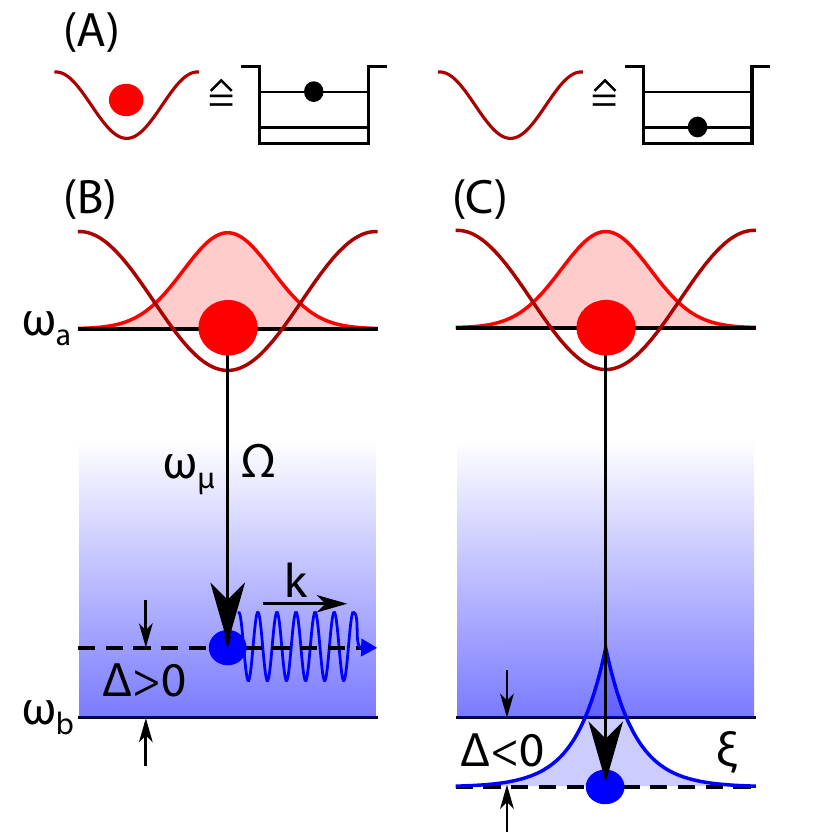}
    \caption{Decay mechanism for an atom confined to a well of a deep, state-selective lattice potential with coupling to a continuum of momentum modes through an internal state transition. (A) The population in the well can be viewed as an occupational spin represented by an excited state (contain 1 atom) or ground state (contain 0 atoms). The coupling, with strength given by the Rabi frequency $\Omega$ and frequency $\omega_\mu$, can be tuned to positive (B) or negative (C) detunings $\Delta$ around the (zero-momentum) boundary of the mode continuum. In (B), $k$ is the momentum of the resonantly coupled freely propagating mode. In (C), an evanescent matter wave with decay length $\xi$ is formed.}
    \label{FIG:setup}
\end{figure}

We consider the experimental situation \cite{deVega2008} of  an atom in a tightly-confining well of a deep optical lattice (with negligible tunneling to other wells) that is coupled to unconfined states via a near-resonant coupling field  of frequency $\omega_\mu$, 
cf. figure \ref{FIG:setup}(A). Such a system may e.g. be realized by using a pair of nondegenerate alkali-atom hyperfine ground states in a state-selective optical potential \cite{Deutsch1998, Jaksch1999, Gadway2010} exposed to radiofrequency or microwave radiation. 
The trapped atom, in the internal state $\ket{a}$, is assumed to be in the harmonic-oscillator ground state with energy $\hbar\omega_a = \hbar\omega_a^0 + d\,\hbar\omega_0/2$ where $\hbar\omega_a^0$ is the bare (untrapped) energy of $\ket{a}$, and $d$ is the system dimensionality (we will set $d=1$ after deriving the Hamiltonian in full generality.) 
The wavefunction in the well is Gaussian,
\begin{equation}
\label{eq:wannier}
\phi_0(\vec{r}) = \frac{1}{\pi^{d/4}a_{ho}^{d/2}}\exp\left[\frac{-\vec{r}^2}{2 a_{ho}^2}\right]
\end{equation} 
with $a_{ho} = \sqrt{\hbar/m\omega_0}$ the harmonic oscillator length. 

The atoms in the unconfined state are assumed to be simple plane waves $\psi_{\vec{k}}(\vec{r})=\exp(i\vec{k}\cdot\vec{r})/L^{d/2}$, having momentum $\vec{k}$ and kinetic energy $E_k = \hbar^2k^2/2m$, where $L\rightarrow\infty$. These states sit on an energy floor associated with the internal energy of $\ket{b}$, namely $\hbar\omega_b=\hbar\omega_b^0$. The system Hamiltonian can be written as
\begin{equation}
\label{eq:fullatomicH}
\hat{H} = \sum_{j=a,b}\int d^dr\hat{\Psi}_j^\dagger(\vec{r})(H_j+\hbar\omega_j^0)\hat{\Psi}_j(\vec{r})+\hat{H}_{ab}
\end{equation}
with the interaction
\begin{equation}
\label{eq:interactionh}
\hat{H}_{ab} = \frac{\hbar\Omega_1}{2}\int d^d r e^{-i\omega_\mu t}\hat{\Psi}_a(\vec{r})\hat{\Psi}_b^\dagger(\vec{r})\hat{\mu}(\vec{r})+H.c.
\end{equation}
The operator $\hat{\mu}(\vec{r})$ annihilates a coupling field quantum. Assuming the coupling field to be classical, $\hat{\mu}(\vec{r})$ can be replaced by its expectation value $\ev{\hat{\mu}}=\sqrt{N}\approx\sqrt{N+1}=\ev{\hat{\mu}^\dagger}$, and can be pulled out of the integral to re-scale the single-photon Rabi frequency to the $N$-photon Rabi frequency, $\Omega_N=\sqrt{N}\Omega_1\equiv\Omega$.

To proceed, the field operators in \eqref{eq:interactionh} are expanded in the basis of states discussed above:
\begin{align}
\hat{\Psi}_a(\vec{r}) &= \phi_0(\vec{r})e^{-i\omega_a t}\hat{a}\\
\hat{\Psi}_b(\vec{r}) &= \sum_{\vec{k}} \psi_{\vec{k}}(\vec{r})e^{-i(E_k/\hbar+\omega_b)t}\hat{b}_{\vec{k}}
\end{align}
The bare-Hamiltonian terms do not contribute to the Schr\"odinger equation governing the time evolution of the state amplitudes in the interaction picture. Hence, the interaction Hamiltonian becomes
\begin{align}
\hat{H}_{ab} &= \frac{\hbar\Omega}{2}\sum_{\vec{k}} \exp[-i(\omega_\mu+\omega_a) t+i(E_k/\hbar+\omega_b)t]\hat{a}\hat{b}_{\vec{k}}^\dagger\nonumber\\
&\qquad\times\int d^d r \phi_0(\vec{r})\psi_{\vec{k}}^*(\vec{r}) + H.c.\\
&= \sum_{\vec{k}} \frac{\hbar\Omega}{2}e^{-i\Delta_k t}\gamma_{\vec{k}}\hat{a}\hat{b}_{\vec{k}}^\dagger+ H.c.
\end{align}
where $\Delta_k$ is a $k$-dependent detuning and $\gamma_{\vec{k}}$ is a Franck-Condon overlap:
\begin{align}
\label{eq:fcdef}
\Delta_k &= \frac{\hbar k^2}{2m}-\Delta; \quad \Delta = \omega_\mu - (\omega_b-\omega_a)\\
\label{eq:fcdefb}
\gamma_{\vec{k}} &= \int d^d r \phi_0(\vec{r})\psi_{\vec{k}}^*(\vec{r})
\end{align}
The atom can either be in the trapped state with no freely propagating modes occupied, $\ket{1_a,\{0\}_{\vec{k}}}$, or it can be in the untrapped state with a freely propagating excitation present, $\ket{0_a,1_{\vec{k}}}$, and hence we represent $\hat{a} = \ket{0_a}\bra{1_a}$.  The interaction Hamiltonian is then reduced to the form
\begin{equation}
\label{eq:intermediatehab}
\hat{H}_{ab} = \sum_{\vec{k}}\frac{\hbar\Omega}{2}\gamma_{\vec{k}}e^{-i\Delta_k t}\hat{b}^\dagger_{\vec{k}}\ket{0_a}\bra{1_a}+H.c.
\end{equation}
By identifying $\ket{0_a}$ and $\ket{1_a}$ with the ground and excited states of an occupational spin as defined in fig. \ref{FIG:setup}A, and upon introducing $g_{\vec{k}} = \Omega\gamma_{\vec{k}}/2$, \eqref{eq:intermediatehab} exactly reproduces the standard Weisskopf-Wigner Hamiltonian for the description of spontaneous photon emission from a two-level atom \cite{Meystre2007} in the interaction picture, 
albeit with different momentum dependences in the $g_{\vec{k}}$ and $\Delta_k$ terms. The differences are caused by the quadratic dispersion relation of the matter-waves, which coincides with that of light in a PBG material \cite{Kofman1994}. 
We note that, while the coupling to the continuum requires the introduction of the external drive, this does not affect the structure of the Hamiltonian, in which the coupling is a simple constant in either case. 

To proceed, we follow the usual approach of expanding the initial state as \cite{Meystre2007}
\begin{equation}
\ket{\Psi(t)} = A(t)\ket{1_a,\{0\}_{\vec{k}}}+\sum_{\vec{k}}B_{\vec{k}}(t)\ket{0_a,1_{\vec{k}}}
\end{equation}
where, due to the choice of the interaction picture, the dynamical phases have been left in the Hamiltonian.  
Application of the Hamiltonian \eqref{eq:intermediatehab} to this initial state, left multiplication by $\bra{1_a,\{0\}_{\vec{k}}}$ or $\bra{0_a,1_{\vec{k}}}$, and cancellation of terms arising from the bare Hamiltonian, then results in the following system of differential equations for the state amplitudes:
\begin{align}
\label{eq:populationdiffeqsa}
\dot{A}(t) &= i\sum_kg^*_{\vec{k}}e^{-i\Delta_kt}B_{\vec{k}}(t) \\
\label{eq:populationdiffeqsb}
\dot{B}_{\vec{k}}(t) &= i g_{\vec{k}}e^{i\Delta_kt}A(t)
\end{align}
The dimension $d$ of the problem changes the Franck-Condon factor, but with this caveat, the preceding derivation is general for any dimensionality. In the following, we restrict the discussion to the one-dimensional case, $d=1$.

\section{Population dynamics}
\label{SEC:markovcase}

To determine the time-dependent amplitude $A(t)$ of the population trapped in the well, we proceed in analogy to the treatment of an excited atom coupled to a PBG material \cite{Kofman1994}. In our case, the excited (ground) state of the emitting atom is replaced by the occupational spin in the well.
First, note that a simple computation shows that
\begin{equation}
\label{eq:fcfullexpression}
\gamma_k = \sqrt{\frac{2\pi^{1/2}a_{ho}}{L}}\exp\left[-\frac{1}{2}k^2a_{ho}^2\right]
\end{equation}
Next, \eqref{eq:populationdiffeqsb} is formally integrated and inserted into \eqref{eq:populationdiffeqsa}. With the assumption that the momenta are closely spaced, i.e. that $L$ diverges,  the sum in \eqref{eq:populationdiffeqsa} is replaced by an integral, and the result for the excited-state amplitude with \eqref{eq:fcfullexpression} is
\begin{equation}
\label{eq:beforemarkovapprox}
\dot{A}(t) = -\frac{a_{ho}(\Omega/2)^2}{\sqrt{\pi}}\int_{-\infty}^\infty dk\int_0^t dt'e^{-k^2a_{ho}^2}e^{i\Delta_k(t-t')}A(t')
\end{equation}
The $k$ integration is carried out first in closed form, as it is simply a Gaussian integral, giving
\begin{equation}
\dot{A}(t) = -\sqrt{2}\int_0^t dt'A(t')\frac{(\Omega/2)^2 e^{i\Delta(t-t')}}{\sqrt{2+i\omega_0(t-t')}}
\end{equation}
which can also be re-written in the form
\begin{equation}
\label{eq:nonmarkov}
\dot{A}(t) = -\int_0^t dt'A(t')G_{1D}(t-t')
\end{equation}
with the correlation function of the continuum \cite{Navarrete_Benlloch2011} (bath correlation function)
\begin{equation}
G_{1D}(\tau) = \frac{(\Omega/2)^2}{\sqrt{1+i\omega_0\tau/2}}e^{i\Delta\tau}
\end{equation}
Eq. \eqref{eq:nonmarkov} is easily solved by the Laplace transform method. Denoting by $\tilde{f}(s)$ the Laplace transform of a function $f(t)$, $\tilde{f}(s) = \mathcal{L}\{f(t)\}$, and using its standard relations (especially the convolution property), one finds
\begin{equation}
\label{eq:laplacesoln}
\tilde{A}(s) = \frac{1}{s+\tilde{G}_{1D}(s)}
\end{equation}
This equation is formally solved by inversion:
\begin{align}
\label{eq:inverselaplace}
e^{-i\Delta t}A(t) &= \frac{1}{2\pi i}\int_{\epsilon-i\infty}^{\epsilon+i\infty}ds \tilde{A}(s+i\Delta)e^{st}\nonumber\\ 
&= \frac{1}{2\pi i}\int_{\epsilon-i\infty}^{\epsilon+i\infty}ds\frac{e^{st}}{s+i\Delta+\tilde{G}_{1D}(s+i\Delta)}
\end{align}
where $\epsilon$ is arbitrarily chosen so that all of the poles of the integrand lie to the left of the integration contour, a vertical line (Bromwich contour).

In general, eq. \eqref{eq:inverselaplace} cannot be solved analytically, but progress can be made, as in the 3D case \cite{Navarrete_Benlloch2011}, by making the assumption of strong coupling $\omega_0\gg s,\Delta$, or equivalently $|\eta|= |(\Delta+is)/\omega_0|\ll1$. This assumption does not accurately capture times below $\omega_0^{-1}$, so the model is expected to break down at very short times. The Laplace transform of the bath correlation function (written in terms of the previously defined $\eta$) is
\begin{equation}
\label{eq:gtilde}
\tilde{G}_{1D}(\eta) = -\sqrt{2\pi}\frac{\Omega^2}{4\omega_0}i\eta^{-1/2}e^{-2\eta}\left(i+\text{Erfi}(\sqrt{2\eta})\right)
\end{equation}
where Erfi is the imaginary error function (c.f. appendix). In strong coupling to leading order in $\eta$ (keeping only the constant and negative power terms), this becomes
\begin{equation}
\label{eq:strongcoupling}
\tilde{G}_{1D}(s) \approx -i\delta_L+\frac{C(1-i)}{\sqrt{s-i\Delta}};\,\, \delta_L = \frac{\Omega^2}{\omega_0};\,\, C = \frac{\sqrt{\pi}}{4}\frac{\Omega^2}{\sqrt{\omega_0}}
\end{equation}
and therefore one must evaluate
\begin{equation}
\label{eq:invlapsc}
e^{-i\Delta t}A(t) = \frac{1}{2\pi i}\int_{\epsilon-i\infty}^{\epsilon+i\infty}ds\frac{e^{st}}{s+i\tilde{\Delta}+C(1-i)/\sqrt{s}}
\end{equation}
The quantity $\delta_L=\Omega^2/\omega_0$ in \eqref{eq:strongcoupling} corresponds to a Lamb shift (c.f. appendix) of the detuning to
\begin{equation}
\tilde{\Delta}= \Delta-\delta_L = \Delta-\Omega^2/\omega_0
\end{equation}

The inversion of the Laplace transform \eqref{eq:invlapsc} is now straightforward, giving 
\begin{align}
\label{eq:Kofmansol}
A(t) &= \exp(i \Delta t)\left[\sum_j\frac{2u_j^2}{3u_j^2+\tilde{\Delta}}\exp(iu_j^2t)\right.\nonumber\\
 &\quad+\left.e^{-i\pi/4}\frac{D}{\pi}\int_0^\infty\frac{\zeta^{1/2}\exp(-\zeta t)d\zeta}{\zeta^3-2i\tilde{\Delta}\zeta^2-\tilde{\Delta}^2\zeta-iD^2} \right]
\end{align}
In this system, $D=\sqrt{2}C$, and $D^{-2/3}$ gives the characteristic time for the dynamics of the model. 
The $u_j$'s are the roots of 
\begin{equation}
\label{eq:rootcubic}
u^3+\tilde{\Delta}u-D=0
\end{equation}
such that $-3\pi/4<$ arg$(u_j)<\pi/4$ to ensure that the roots $u_j$ do not lie on the branch cut in $s$-space. 

\begin{figure}[h!]
\centering
    \includegraphics[width=0.95\columnwidth]{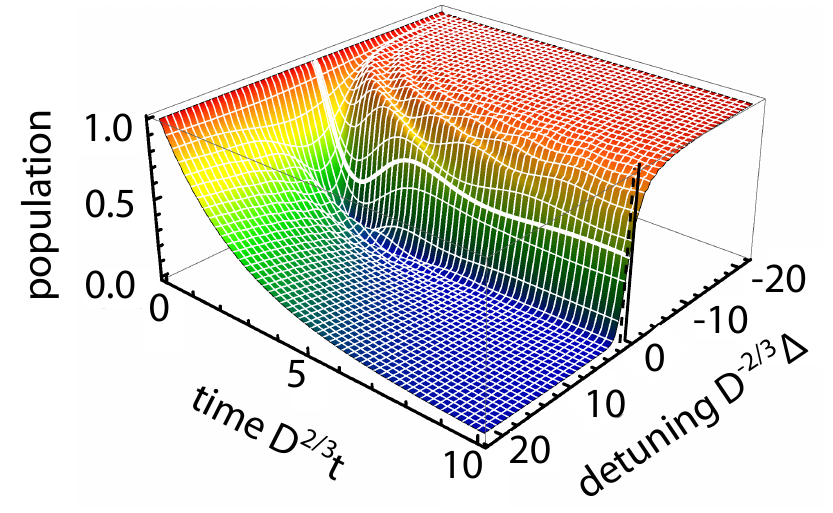}
    \caption{Computed population $|A(t)|^2$ in the potential well as a function of drive time and detuning in characteristic time and frequency units. The thick white curve corresponds to $\Delta=0$ (see vertical solid line), and the Lamb shift is $D^{-2/3}\delta_L  = 0.1$ (see the dotted vertical line on the detuning axis). The chosen frequency scale is $D^{2/3}/2\pi\approx 200$ Hz. For the depicted plot, $\omega_0$ and $\Omega$ are fixed by matching to experimentally reasonable values ($\omega_0\approx 2\pi\times30$ kHz and $\Omega/\omega_0 = 0.03$).}
   \label{FIG:fullinformation}
\end{figure}

Since \eqref{eq:Kofmansol} is analytic, it can be evaluated over any range of interest; an example with the salient features is shown in fig. \ref{FIG:fullinformation}. In analogy to what is observed in PBG materials near a band gap \cite{Kofman1994,Lambropoulos2000}, depending on the detuning there is a transition from a (nearly) exponential decay to oscillatory behavior, in which the atomic population decays slightly but remains trapped in the well.

For large positive detunings $\Delta\gg0$, the system is so far away from the continuum boundary that the density of levels looks essentially unchanged from the original situation considered by Weisskopf and Wigner. In this regime, one may make the standard Markov approximation, the details and results of which can be found in the appendix. In particular, the population is found to decay exponentially with a rate $\Gamma = \delta_L\sqrt{\omega_0/\Delta}\exp(-2\Delta/\omega_0)$.

For that are smaller than the Lamb shift, $\Delta<\delta_L$ (this includes arbitrarily large negative detunings), the population in the well does not decay completely, even for arbitrarily long times. This is a consequence of the fact that eq. \eqref{eq:rootcubic} has a real root for $\tilde{\Delta}\lesssim0$, i.e. for $\Delta\lesssim\delta_L$, which signals that the solution has an imaginary pole at $s = -\text{i}u_j^2$, corresponding to a long-lived excitation. This excitation is a dressed state which evolves at the frequency determined by the real root of eq.\eqref{eq:rootcubic}. [For more details, see sections \ref{SEC:boundpairs} and \ref{SEC:analogapbs}]. 
The behavior for $\Delta=0$, i.e. the resonant case, is reminiscent of a damped Rabi oscillation, though the frequency appears to vary weakly with time. (For much stronger couplings than considered here, the trap potential and the flat continuum hybridize into dressed potentials, eventually leading to an undamped Rabi oscillation \cite{Reeves2015}). We note that the case under consideration in fig. \ref{FIG:fullinformation} corresponds roughly to the small coupling limit considered in \cite{Reeves2015}, however with an initial state having a broad momentum spread. 

In contrast to 3D results discussed in \cite{deVega2008}, the oscillations in the population in the 1D case persist for many characteristic times at negative detunings. Furthermore, the decay rate is maximum at detunings around $\delta_L$, and it becomes slower as the detuning is increased. The pronounced oscillatory behavior in 1D is consistent with the divergence of the 1D density of states at zero energy. 
While the fundamental differences between 1D and 3D systems would be difficult to measure in PBG materials \cite{Tocci1996} (see also \cite{Lodahl2004}), the tunability of the atom-optical system makes it an ideal candidate for the exploration of dimensional effects. 

\section{Momentum Distributions of Emitted Matter Waves}
\label{SEC:momentumdistributions}

The coupled differential equations \eqref{eq:populationdiffeqsa} and \eqref{eq:populationdiffeqsb} contain the amplitudes of both the occupied well and the emitted matter waves. Using eq \eqref{eq:Kofmansol} for $A(t)$, we may formally integrate the eq. \eqref{eq:populationdiffeqsb} for $\dot{B}_k(t)$, leading to
\begin{equation}
\label{eq:momentumcompB}
B_k(t) = i\frac{\Omega}{2}\sqrt{\frac{2\pi^{1/2}a_{ho}}{L}}e^{-k^2a_{ho}^2/2}\int_0^t e^{i\Delta_k t'}A(t')dt'
\end{equation}
The absolute square of \eqref{eq:momentumcompB} gives the time-dependent momentum distribution:
\begin{align}
\label{eq:momentumdistrib}
L|B_k(t)|^2 &= \sqrt{\pi}\frac{\Omega^2}{2}a_{ho}e^{-k^2a_{ho}^2}I(k,t)\nonumber\\
I(k,t) &=\left|\int_0^t \exp\left[i\left(\frac{\hbar k^2}{2m}-\Delta \right) t' \right]A(t')dt' \right|^2
\end{align}
which for long times defines the emission spectrum \footnote{In the formalism of \cite{Berman2010B}, this matter wave system corresponds to $\gamma=\sqrt{2\pi}\Omega^2/\omega_0$ and $F(\omega_k)=\exp(-2\omega_k/\omega_0)\sqrt{\omega_0/\omega_k}\Theta(\omega_k)$, a spectral weight function not previously considered.}
$S(\omega_k)  = \lim_{t\rightarrow\infty}|B_k(t)|^2$.

Using the numerical solution for $A(t)$ with parameters $0\leq t'\leq t$, $\Delta$, $\Omega$, and $\omega_0$, the integral $I(k,t)$ and the momentum distribution $L|B_k(t)|^2$ can then be computed numerically.

\begin{figure}[h!]
\centering
    \includegraphics[width=0.95\columnwidth]{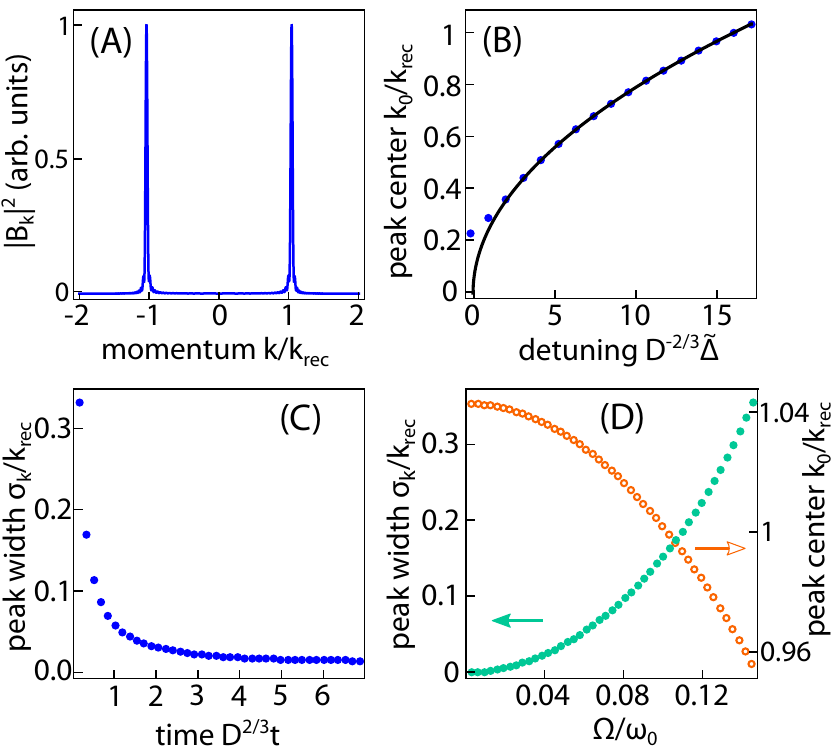}
    \caption{Momentum distributions of emitted matter waves (positive detunings, Markovian limit). (A) Sample momentum distribution taken at time $D^{2/3}t\approx 1.5$ and positive detuning $D^{-2/3}\tilde{\Delta}\approx 17$. The width and position of the peaks depends on time and detuning, but the shape is representative for a large range of parameters. (B) Calculated momentum peak position $k_0$ (in units of $k_{rec}$, see text) as a function of detuning $D^{-2/3}\tilde{\Delta}$ for $D^{2/3}t\approx 4.0$. The solid curve is a guide to the eye for the expected momentum based on exact energy conservation (see text). (C)  Momentum peak width $\sigma_k$ vs. time for $D^{-2/3}\tilde{\Delta}\approx17$, saturating at a value set by the decay rate (see text). (D) Momentum peak width (green, filled circles, left vertical axis) and momentum peak position (orange, unfilled circles, right vertical axis) vs. Rabi frequency $\Omega$ for $D^{2/3}t\approx1.5$ and $D^{-2/3}\tilde{\Delta}\approx17$. 
}
    \label{FIG:momentumwaterfall}
\end{figure}

An example for the Markovian limit is shown in fig. \ref{FIG:momentumwaterfall}A. The doubly-peaked momentum structure is characteristic of a system in 1D with left-right symmetry; while the individual peaks are not generally symmetric around their centers, their location varies with detuning in a simple way. The detuning supplies a kinetic energy $\hbar\tilde{\Delta}$ to the transferred atoms, which corresponds to a particular momentum $k(E_{kin})/k_{rec} = \sqrt{\hbar\Delta/\hbar\omega_{rec}}$, where we have introduced $\hbar\omega_{rec} = (\hbar k_{rec})^2/2m$ with $k_{rec} = 2\pi/\lambda_{latt}$ and $\lambda_{latt}$ is the wavelength of the optical lattice. Fig. \ref{FIG:momentumwaterfall}B shows the extracted peak location, $k_0 = \sqrt{2mE_0}/\hbar$, where $E_0$ is the computed peak energy in fig. \ref{FIG:momentumwaterfall}A from the numerical data, as well as a no-free-parameter fit to the data of square-root form. The close agreement with the simulated data 
degrades at small detunings, where the doubly-peaked structure is lost due to edge effects near the boundary, such that one characteristic momentum in this regime cannot be identified. 

It is also seen that the momentum distribution evolves in time, starting from wide peaks and tending to a tightly confined value, cf. \ref{FIG:momentumwaterfall}C. To understand this behavior, consider that in the Markovian regime, the decay is exponential, and so will have an (approximately) Lorentzian emission spectrum $S(\omega)$ \cite{Berman2010B}. The width of the momentum distribution $\sigma_k = \sigma_\omega/2k_0$ (where $\sigma_\omega$ is the width from a Lorentzian fit to the energy distribution) is limited at early times by the corresponding spectral Fourier width ($\Delta t\Delta\omega\approx1$), but at long times should tend to a finite value set by the line width $\Gamma$ of the excited state, given by \eqref{eq:definegamma}. Indeed, computation of the emission spectrum $S(\omega)$ (not shown) by an appropriate change of variables, and extracting its width yields a value of 1.25$\Gamma$ for long times, in good qualitative agreement with the expectation.

So far, the discussion of fig. \ref{FIG:momentumwaterfall} has focused on a fixed Rabi frequency $\Omega$. When $\Omega$ is varied, both the decay rate $\Gamma$ and the Lamb shift $\delta_L$ change proportional to $\Omega^2$. Thus, the width in kinetic energy at long times varies quadratically with $\Omega$. Furthermore, the extracted peak separation decreases with increasing Rabi frequency as the growing Lamb shift $\delta_L$ pulls the lattice-trapped state to a lower energy. 

\section{Evanescent matter-waves}
\label{SEC:boundpairs}

As already explained in section \ref{SEC:momentumdistributions}, the incomplete decay of the population in the well for some detunings is due to the creation of a stable excitation. When the Laplace transforms of \eqref{eq:populationdiffeqsa} and \eqref{eq:populationdiffeqsb} contain an imaginary pole (relating to a real $u_j$ in eq.\eqref{eq:rootcubic}), the corresponding time domain behavior features a coherent superposition of freely-propagating modes that evolve not with their own dynamical phases, but with the phase determined by the purely imaginary pole, denoted here by $\omega_B$. The resulting wavefunction corresponding to this sum can be shown to be (see section \ref{SEC:analogapbs})
\begin{equation}
\label{eq:boundpairwavefunction}
\psi_{B}^{mw}(x) = c_B^{1/2}\int_0^\infty\frac{g^*(\omega)}{\omega-\omega_{B}}\rho(\omega)\ip{x}{0_a,1_\omega}d\omega
\end{equation}
where a switch to an energy (rather than momentum) representation has been made upon introducing the energy density of levels $\rho(\omega)$. The frequency $\omega_{B}$ is determined by the relationship \cite{Kofman1994}
\begin{equation}
\label{eq:omegabp}
\omega_{B} = \Delta+\int_0^\infty\frac{G(\omega)}{\omega_{B}-\omega}d\omega
\end{equation}
where $G(\omega) = |g(\omega)|^2\rho(\omega)$. For the system considered here, $|g(\omega)|^2$ and $\rho(\omega)$ are given by
\begin{equation}
\label{eq:bathfuncs}
|g(\omega)|^2 = \frac{\sqrt{\pi}a_{ho}}{L}\frac{\Omega^2}{2}e^{-2\omega/\omega_0};\,\, \rho(\omega) = \frac{L}{\pi}\sqrt{\frac{2m}{\hbar\omega}}
\end{equation}

\begin{figure}[h!]
\centering
    \includegraphics[width=0.95\columnwidth]{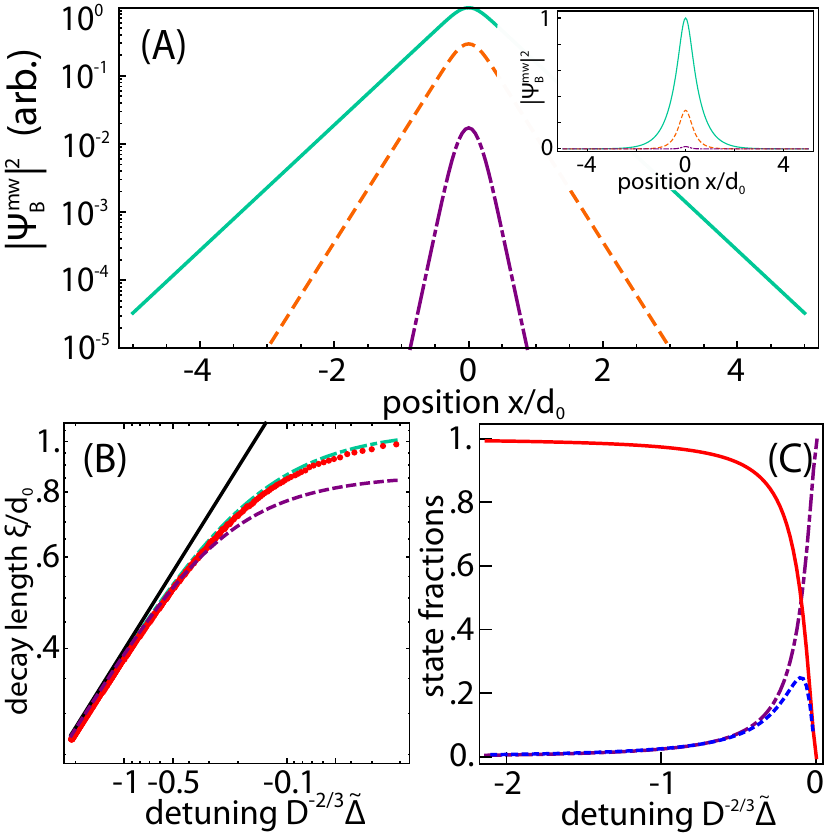}
    \caption{Characteristics of the evanescent matter-wave. (A) Density distribution as a function of distance in lattice spacings $d_0 = \lambda_{latt}/2$ for $D^{-2/3}\tilde{\Delta} =$ -27 (purple, dot-dashed), -2.7 (orange, dashed), and -.27 (green, solid). (B)  Extracted decay length in lattice spacings as a function of the detuning (red dots). The slope of the black line is $-1/2$ (after reversing the horizontal axes), the scaling expected in 1D for large detunings. The purple (dotted) curve is the model based on eq. \ref{eq:equivstarkshift}, and the green (dot-dashed) curve is a more sophisticated model (see text). (C) The relative populations of the well (i.e. excited state of the well) (red, solid), and the evanescent matter wave (blue, dashed) as a function of detuning. We also depict the population that is radiated away (purple, dot-dashed).}
   \label{FIG:boundpairs}
\end{figure}

Integration of \eqref{eq:omegabp} with \eqref{eq:bathfuncs} leads to the analytic result
\begin{equation}
\label{eq:fullomegab}
\omega_B = \Delta-\sqrt{\frac{\pi}{2}}\delta_L\sqrt{\frac{\omega_0}{|\omega_B|}}\exp\left(\frac{2|\omega_B|}{\omega_0}\right)\text{Erfc}\left(\sqrt{\frac{2|\omega_B|}{\omega_0}}\right)
\end{equation}
In this expression, Erfc($z$) is the complement of the error function, i.e. $\text{Erfc}(z) = 1-\text{Erf}(z)$. The assumption of strong lattice confinement, $\omega_0$ being much larger than any other frequency scale in the problem, allows for the Taylor expansion of this result, and keeping the leading order term $\propto \sqrt{\omega_0/\omega_B}$ and the sub-leading order term which does not depend on $\omega_B$, results in
\begin{equation}
\label{eq:polecomparison}
\omega_B = (\Delta-\delta_L)-\frac{i D}{\sqrt{\omega_B}}
\end{equation}
Comparison with the denominator in eq.\eqref{eq:invlapsc} reveals this to be the same form upon rotating $s=-\text{i}\omega_B$. Thus the dynamical equations in sec. \ref{SEC:markovcase} have been recovered in a slightly different formalism. 

The matter-wave of eq. \eqref{eq:boundpairwavefunction} contains contributions from all possible eigenstates $\ip{x}{0,1_\omega}$ with frequency $\omega$. Using $k(\omega) = \sqrt{2m\omega/\hbar}$, the eigenstates are $\varphi_\omega(x) = \ip{x}{0_a,1_\omega} \propto \cos[k(\omega)x]$. In this way, the spatial profile of the wave can be constructed by solving for $\omega_{B}$ given at fixed detuning, and then computing \eqref{eq:boundpairwavefunction}, c.f. fig. \ref{FIG:boundpairs}A.

As is particularly apparent from fig. \ref{FIG:boundpairs}A, $\psi_B(x)$ is exponentially localized. Note that the decay is different from the 3D case, where a Yukawa-type profile is seen to arise away from the center of the bound state \cite{deVega2008}. Fitting an exponential to the wings of the computed states and extracting this decay for a large range of negative detunings allows for a comparison of the decay length to an evanescent-wave model of the form
\begin{equation}
\label{eq:xiansatz}
\xi = a_{ho}\sqrt{\frac{\omega_0}{2|\tilde{\Delta}|}}
\end{equation}
The resulting fit decay lengths are shown in fig. \ref{FIG:boundpairs}B (red circles) [We plot three sample curves in units of $d_0=\lambda_{latt}/2$ with $\lambda_{latt} = 792.5$ nm the lattice wavelength as the characteristic length scale of the problem. We do so looking ahead to an experimental implementation, c.f. sec \ref{SEC:conclusion}.] It should be noted that eq. \eqref{eq:xiansatz} matches the expectation in which one na\"ively assumes a matter wave with negative energy and a corresponding imaginary wave-vector $\kappa$ set by the detuning only for large negative detunings $\tilde{\Delta}$.  To properly capture the physics of the system, $\hbar\omega_B$ (the bound-state energy, see sec. \ref{SEC:analogapbs}) must be used instead of the detuning, with the associated characteristic length
\begin{equation}
\label{eq:properxi}
\xi=a_{ho}\sqrt{\frac{\omega_0}{2|\omega_B(\Delta)|}}
\end{equation}
To determine $\omega_B(\Delta)$, the soft cutoff in the integral in eq.\eqref{eq:omegabp}, $\exp(-2\omega/\omega_0)$ can be approximated by a sharp cutoff of the integral at $\omega\approx\omega_0$. Expanding the integrand to leading order in the, assumed small, quantity $\omega/\omega_0$ and integrating, leads to an effectively quadratic equation for $\omega_B$. This equation has a non-trivial negative root 
\begin{equation}
\label{eq:equivstarkshift}
\omega_B = \frac{\Delta}{2}+\frac{1}{2}\sqrt{\Delta^2+\Xi\Omega^2}
\end{equation}
which for small $\Omega$ coincides with the set detuning from the edge of the continuum. The second term has the form of a generalized Rabi frequency $\sqrt{\Delta^2+\Xi\Omega^2}$ ($\Xi$ is a numerical prefactor of order 1), and suggests that for intermediate detunings, the correction to the decay length is due to an AC Stark shift induced by the coupling (note that it is independent of the lattice potential). This expression \eqref{eq:properxi} with \eqref{eq:equivstarkshift}, however, underestimates the decay length for small detunings, c.f. fig. \ref{FIG:boundpairs}B (purple, dotted). In order to recover the proper behavior in this limit, eq. \eqref{eq:fullomegab} must be solved. For the regimes of validity of this work, this amounts to solving eq. \eqref{eq:polecomparison}, an effectively \textit{cubic} equation for the bound state energy, whose real root, when inserted into eq. \eqref{eq:properxi}, gives the true decay length of the system. This curve is also shown in fig. \ref{FIG:boundpairs}B (green, dot-dashed), giving much better agreement with the fit decay lengths.

\section{Analogy to atom-photon bound states in PBG materials}
\label{SEC:analogapbs}

The exponentially localized matter-wave discussed in the previous section is part of a larger state also comprising a superposition of the occupational spin state of the well. This is in direct analogy to the so-called \textit{atom-photon bound state} in PBG materials \cite{Bykov1975, John1984, John1990, John1991, Kofman1994, Lambropoulos2000} as a ``dynamic state in a superposition of the excited and ground states with an admixture of a photon ``cloud'', which surrounds the atom''\cite[p. 866]{Bykov1975}.

In order to elucidate the composition of our ``lattice well-atomic matter-wave bound state'', note that the Laplace transforms of \eqref{eq:populationdiffeqsa} and \eqref{eq:populationdiffeqsb} can be shown to be:
\begin{align}
\label{eq:kofmanlaplacea}
\tilde{A}(s+i\Delta) &= [s+i\Delta+J(s)]^{-1}\\
\label{eq:kofmanlaplaceb}
\tilde{B}_\omega(s) &= \frac{-ig^*(\omega)\tilde{A}(s+i\Delta)}{s+i\omega}
\end{align}
where
\begin{equation}
\label{eq:jdefinition}
J(s) = i\delta_L+\alpha(1-i)\sqrt{\pi/s}
\end{equation}
is the approximate form of the Laplace transform of the bath-memory kernel in the limit of strong coupling.

Suppose that \eqref{eq:kofmanlaplacea} has an imaginary pole at $s=-i\omega_B$. Then inverting the Laplace transform $\tilde{A}(s)$ will contain this pole within the inversion contour, and so schematically,
\begin{equation}
\label{eq:kofmanat}
A(t) = \frac{1}{2\pi i}\int_{\zeta-i\infty}^{\zeta+i\infty}\tilde{A}(s)e^{st}ds = \text{Res}_{-i\omega_B}\left[\tilde{A}(s)e^{st}\right] + A_c(t)
\end{equation}
where the residue may be computed as
\begin{align}
\label{eq:residuecomp}
\text{Res}_{-i\omega_B}\left[\tilde{A}(s)e^{st}\right] &= \left.\frac{e^{st}}{\bd_s(s+i\Delta+J(s))}\right|_{s=-i\omega_B}\nonumber\\
&=\left(1+\bd_sJ(s)|_{s=-i\omega_B} \right)^{-1}e^{-i\omega_Bt} \nonumber\\
&=c_Be^{-i\omega_Bt}
\end{align}
Thus \eqref{eq:kofmanat} takes the form
\begin{equation}
\label{eq:kofmanat2}
A(t) = c_Be^{-i\omega_Bt}+A_c(t)
\end{equation}
The form of $A_c(t)$ is model dependent, and it cannot be written in analytic form for the model we consider. (It roughly corresponds to the integral term in equation \eqref{eq:Kofmansol}, along with decaying pieces of the summation term.) The salient features, specifically that its modulus tends to a constant value less than one, are however independent of the specific form.

The result from eq. \eqref{eq:kofmanat2} is directly inserted into the equation for $\tilde{B}_\omega(s)$ and, defining $\tilde{\gamma}(s) = (s+i\omega)^{-1}$, using the convolution property of Laplace transforms to solve for $B_\omega(t)$:
\begin{align}
\label{eq:kofmanbt}
\tilde{B}_\omega(s) &= -ig^*(\omega)\tilde{A}(s)\tilde{\gamma}(s) \nonumber\\
\Rightarrow B_\omega(t) &= -ig^*(\omega)\int_0^t\gamma(t-\tau)A(\tau)d\tau
\end{align}
The convolution in \eqref{eq:kofmanbt} can be evaluated
\begin{align}
\label{eq:kofmanbt2}
B_\omega(t) &= -ig^*(\omega)\int_0^t e^{-i\omega(t-\tau)}(c_Be^{-i\omega_B\tau}+A_c(t))d\tau\nonumber\\
&= \frac{-ig^*(\omega)ic_B(e^{-i\omega t}-e^{-i\omega_Bt})}{\omega-\omega_B}+B'_{\omega,c}(t) \nonumber\\
&= \frac{-g^*(\omega)c_Be^{-i\omega_Bt}}{\omega-\omega_B}+B_{\omega,c}(t)
\end{align}
Again, the exact form of $B_{\omega,c}(t)$ has no closed form. Now, combining \eqref{eq:kofmanat2} and \eqref{eq:kofmanbt2} into the state expansion for  the system yields the state
\begin{align}
\label{eq:boundpairform}
\ket{\Psi(t)} &= A(t)\ket{1_a,\{0\}}+\int_0^\infty B_\omega(t)\rho(\omega)\ket{0_a,1_\omega}d\omega \nonumber\\
&= c_Be^{-i\omega_Bt}\ket{1_a,\{0\}}+A_c(t)\ket{1_a,\{0\}}\nonumber\\
&\quad\quad-c_Be^{-i\omega_Bt}\int_0^\infty\frac{g^*(\omega)}{\omega-\omega_B}\rho(\omega)\ket{0_a,1_\omega}d\omega\nonumber\\
&\quad\quad+\int_0^\infty B_{\omega,c}(t)\rho(\omega)\ket{0_a,1_\omega}d\omega\nonumber\\ 
&=\ket{\psi_B}e^{-i\omega_B t}+\ket{\Psi_c(t)}
\end{align}
where
\begin{equation}
\label{eq:psibeq}
\ket{\psi_B} = c_B\left(\ket{1_a,\{0\}}-\int_0^\infty\frac{g^*(\omega)}{\omega-\omega_B}\rho(\omega)\ket{0_a,1_\omega}d\omega\right)
\end{equation}
and 
\begin{equation}
\label{eq:psiceq}
\ket{\Psi_c(t)} = A_c(t)\ket{1_a,\{0\}} + \int_0^\infty B_{\omega,c}(t)\rho(\omega)\ket{0_a,1_\omega}d\omega
\end{equation}
The state $\ket{\psi_B}$ has no time dependence outside of the phase $e^{-i\omega_Bt}$, which we now identify with the energy of the bound state, $\hbar\omega_B$. The first term in eq. \eqref{eq:psibeq} corresponds to the excited state of the lattice well, whereas the second term yields the evanescent wave of eq. \eqref{eq:boundpairwavefunction}, including the ground state of the well. 
The state $\ket{\Psi_c(t)}$ satisfies, 
\begin{equation}
|\Psi_c(t)|^2 = 1-c_B
\end{equation}
because $|\psi_B|^2 = c_B$. Since $c_B$ is the probability to remain bound, the interpretation of $\ket{\Psi_c(t)}$ becomes clear. It represents the atomic population in the well that is released into propagating modes and does not return in time. The existence of this term is understandable if one considers that a sudden turn on of the coupling at $t=0$, as in the preceding treatment, represents a transient, resonant coupling to many different momentum modes. (We expect that this effect should vanish if the coupling is turned on adiabatically.)

We now further analyze the internal-state composition of the state $\ket{\psi_B}$. 
The constant, $c_B$, which depends on the bound state energy $\hbar\omega_B$, has the full form
\begin{align}
\label{eq:cb}
c_B &= \left\{1+\frac{\delta_L}{\sqrt{2\pi}\omega_0}\frac{\exp(2|\nu_B|)}{2|\nu_B|^{3/2}}\left[2\exp(-2|\nu_B|)\sqrt{2\pi|\nu_B|}\right.\right.\nonumber\\
&\quad\quad\quad+\left.\left.\pi(1-4|\nu_B|)\text{Erfc}\left(\sqrt{2|\nu_B|}\right)\right]\right\}^{-1}
\end{align}
where the parameter $\nu_B$ is defined to be $\omega_B/\omega_0$. This constant gives the probability amplitude to measure the system in the lattice-well atomic matter-wave bound state, which as noted, is a superposition of the excited occupational spin state and the evanescent wave. The probability to measure the atom in the  bound state, starts at 0 for a detuning of $\delta_L$ and then increases monotonically with decreasing detuning before saturating at a value of 1, c.f. fig. \ref{FIG:boundpairs}C (red, solid).

The relative proportion of the free space modes which make up 
the evanescent wave in $\ket{\psi_B}$ initially rises with increasing (negative) detuning before reaching a maximum and then dropping off to essentially zero (blue, dotted curve in fig. \ref{FIG:boundpairs}C). For larger and larger negative values of the detuning, the bound state energy $\hbar\omega_B$ sinks farther and farther below the continuum boundary, and in this regime, it is impossible for the free-particle modes to participate in the formation of the bound state in a significant way, leading to the above-noted drop-off behavior.

\section{Conclusion}
\label{SEC:conclusion}

In this work, we have explored in detail spontaneous emission behavior of an atom trapped in a well of a deep optical lattice with variable coupling to free space. The boundary strongly modifies the decay of the population in the lattice well, which displays a crossover from Markovian to non-Markovian dynamics. The emitted matter-wave spectrum at positive detunings is matched well by a simple model for freely-propagating massive particles, and the evanescent wave state formed for negative detunings decays exponentially away from the well. 

The single-particle model discussed in this work may be observed in an experiment using a sparse array of ultracold atoms confined to the ground band of a deep state-dependent optical lattice potential \cite{Deutsch1998, Jaksch1999,Gadway2010}. Specifically, by applying a coupling field of varying strength and detuning from atomic resonance, all regimes discussed in this work can be explored. By using e.g. rubidium-87 atoms in two hyperfine states in a state dependent optical potential at $\lambda_{latt}=792.5$ nm with transverse lattice tube confinement, it is possible to prepare sparse clouds of pinned atomic impurity atoms in the lattice \cite{Gadway2011}. After applying microwave radiation at $\approx 6.8$ GHz driving a hyperfine transition to an untrapped state, the population in the lattice as well as the momentum distribution can be measured in time-of-flight using state sensitive absorption imaging. By varying the exposure time, the evolution of the populations can thus be tracked, and both exponential decay and oscillatory behavior can be extracted \cite{Krinner2016}.

Furthermore, as already discussed for the 3D case \cite{deVega2008}, it should be possible to engineer a Hubbard model  with long range hopping terms 
with a Markovian coupling of
\begin{equation}
\Gamma_{j-l} = \int_0^\infty d\tau G_{j-l}(\tau) = -i\sqrt{2\pi}\delta_L\sqrt{\frac{\omega_0}{|\tilde{\Delta}|}}e^{-|j-l|/\xi}
\end{equation}
where $\xi$ is given by \eqref{eq:xiansatz}, $\delta_L = \Omega^2/\omega_0$ as above, and $j$ and $l$ are lattice site labels. 
Since $\xi$ and $\sqrt{\omega_0/|\tilde{\Delta}|}$ can be tuned independently, this may be used to implement lattice models with long-range tunneling \cite{deVega2008}. With a view towards studying many-body physics, we note that recent theoretical efforts have demonstrated the possibility of creating $N>1$ bound state wavefunctions \cite{Calajo2016, Shi2016}. Furthermore, the AQD model \cite{deVega2008} also provides the basis for studies of superradiant decay, complementing the recent observation of superradiance in waveguide QED systems \cite{Goban2015}. 

\begin{acknowledgments}
We acknowledge funding from the National Science Foundation (Grant Nos. PHY-1205894 and PHY-1607633). M.S. gratefully acknowledges support from a GAANN fellowship from the DoEd. A.P. acknowledges support by a grant from ESPOL-SENESCYT. We thank T. C. Weinacht for stimulating discussions about fundamental aspects of the Weisskopf-Wigner model, and M. G. Cohen for a critical reading of the manuscript.
\end{acknowledgments}

\appendix*
\section{Spontaneous decay in the Markovian regime}
\label{APP:WignerWeisskopf}
We briefly summarize the steps required to arrive at the usual Markov treatment of spontaneous decay in our system, which matches most textbook approaches. We start from \eqref{eq:beforemarkovapprox}.

If $A(t')$ is slowly varying compared to $\exp(i\Delta_k(t-t'))$, $A(t')$ may be replaced by $A(t)$ and removed from the time integral, whose upper limit is formally taken to infinity. These operations amount to making a Markov (or Weisskopf-Wigner) approximation. The time integral can then be performed 
\begin{align}
\label{eq:epsreg}
\int_0^tdt'e^{-i\Delta_k(t-t')} &= \int_0^td\tau e^{-i\Delta_k\tau}\nonumber\\
&\approx \lim_{\veps\rightarrow0}\int_0^\infty d\tau e^{-i\Delta_k\tau-\veps\tau}\nonumber\\
&= \lim_{\veps\rightarrow0}\frac{i}{\Delta_k-i\veps}
\end{align}
where an epsilon regulator is introduced to ensure convergence of the integral; it will be taken to zero at the end of the computation. Therefore, the evolution equation becomes
\begin{equation}
\label{eq:beforechange}
\dot{A}(t) = -\frac{a_{ho}\Omega^2}{4\sqrt{\pi}}A(t)\lim_{\veps\rightarrow0}\int_{-\infty}^\infty dk \frac{ie^{-k^2a_{ho}^2}}{\Delta_k-i\veps}
\end{equation}
This expression can be written as a frequency integral after a change of variables
\begin{equation}
\label{eq:komegajacobian}
\omega_k = \frac{\hbar k^2}{2m}\,\Rightarrow\, d\omega_k=\frac{\hbar}{m}k dk \,\Rightarrow\, a_{ho}dk = \frac{d\omega_k}{\sqrt{2\omega_0\omega_k}}
\end{equation}
Because \eqref{eq:beforechange} is an even function of $k$, with the change of variables \eqref{eq:komegajacobian}, the momentum integral can be written in frequency variables as
\begin{equation}
\label{eq:freqform}
\dot{A} = -A\frac{\delta_L}{\sqrt{2^3\pi}}\lim_{\veps\rightarrow0}\int_0^\infty d\omega_k \frac{i\exp(-2\omega_k/\omega_0)\sqrt{\omega_0/\omega_k}}{\omega-\Delta-i\veps}
\end{equation}
with $\delta_L = \Omega^2/\omega_0$. Taking inspiration from the usual treatment of Weisskopf-Wigner for an atom-light system, both a decay rate and Lamb shift are expected to appear in the solution of \eqref{eq:freqform}. In order to see that a similar phenomenon happens here, recall the Sokhotski-Plemelj theorem, whose precise statement is as follows: Suppose that $a<0<b$, then
\begin{equation}
\label{eq:sokhotskiplemelj}
\lim_{\veps\rightarrow0^+}\int_a^b\frac{f(x)}{x\pm i\veps}dx = \mp i\pi f(0)+\mathcal{P}\int_a^b\frac{f(x)}{x}dx
\end{equation}
where $\mathcal{P}$ indicates the Cauchy principal part of the integrand. Note that if the variable shift  $\omega'=\omega-\delta$ with unit Jacobian is introduced such that the limits of integration in \eqref{eq:freqform} become $-\Delta$ and $\infty$, then for positive detunings $\Delta>0$ only, the hypotheses of the theorem are satisfied, and the integral becomes
\begin{align}
\label{eq:positivedetuning}
\dot{A}(t) &= -A(t)\frac{\delta_L}{\sqrt{2^3\pi}}\left[\pi\sqrt{\frac{\omega_0}{\Delta}}e^{-2\Delta/\omega_0}\right.\nonumber\\
&\quad\left.+i\mathcal{P}\int_0^\infty d\omega_k \frac{e^{-2\omega_k/\omega_0}\sqrt{\omega_0/\omega_k}}{\omega_k-\Delta} \right];\,\,\,\Delta>0
\end{align}
The first term is real, and therefore gives a decay rate. The second term is purely imaginary, and is thus analogous to the Lamb shift. The constant $\delta_L$ from the body of the main text sets the overall scale for the size of both the decay and the shift terms. Note that this second integral is analytically tractable. Furthermore, for $\Delta<0$, the singularity of the integrand is outside the range of integration, and so there is no problem just evaluating the integral by brute force. In this case, the computed integral is purely imaginary. The result is:
\begin{align}
\label{eq:1dcomputation}
\dot{A}(t) &= -A(t)\frac{\delta_L}{2}\left\{e^{-2\Delta/\omega_0}\sqrt{\frac{\pi\omega_0}{2\Delta}}\Theta(\Delta)\right.\nonumber\\
&\quad\left.+ie^{2|\Delta|/\omega_0}\sqrt{\frac{\pi\omega_0}{2|\Delta|}}\left[\text{Erfc}\left(\sqrt{2\frac{|\Delta|}{\omega_0}}\right)\Theta(-\Delta)\right.\right.\nonumber\\
&\qquad\qquad\qquad\qquad \left.\left.+\text{Erfi}\left(\sqrt{2\frac{\Delta}{\omega_0}} \right)\Theta(\Delta)\right]\right\}
\end{align}
 $\Theta(z)$ is the Heaviside step function, Erfc is the complementary error function, 1-Erf($z$), and Erfi is the imaginary error function, $-i$Erf($iz$). 
The population in the excited state evolves in time as $ P_e(t) = |A(t)|^2$ with
\begin{equation}
\label{eq:markovpop}
P_e(t) = \exp\left(-\frac{\Omega^2}{\omega_0}\sqrt{\frac{\pi\omega_0}{2\Delta}}e^{-2\Delta/\omega_0}\Theta(\Delta) \times t\right) 
\end{equation}

Furthermore, as a Markov process, this limit must agree with a simple Fermi's Golden Rule computation. Indeed, by calculating $d|P_e|/dt \equiv \Gamma$, one obtains agreement of the two approaches since
\begin{equation}
\label{eq:definegamma}
\Gamma = \frac{2\pi}{\hbar}|\bra{0_a}\hat{H}_{ab}\ket{1_a}|^2\rho(E=\hbar\Delta)
\end{equation}
with
\begin{align}
\rho(E=\hbar\Delta) &= \frac{L}{\pi\hbar}\sqrt{\frac{m}{2\hbar\omega_0}}\sqrt{\frac{\omega_0}{\Delta}}\Theta(\Delta) \\
|\bra{0_a}\hat{H}_{ab}\ket{1_a}|^2 &= \frac{\hbar^2\Omega^2\pi^{1/2}}{2L}\sqrt{\frac{\hbar}{m \omega_0}}e^{-2\Delta/\omega_0}
\end{align}
where $\hat{H}_{ab}$ is the interaction term in the main text. Thus, the externally applied coupling (whose strength is the Rabi frequency $\Omega$) plays the role of resonant ``vacuum'' fluctuations in this model, which ``stimulate'' a spontaneous decay from the excited state.  We note that, in contrast to the optical case, where the Lamb shift and spontaneous decay rate are fixed by atomic properties, in the atom-optical system they are tunable because $\Omega$ and $\Delta$ are experimental control parameters, yielding a variable ratio $\Gamma/\delta_L = \sqrt{\omega_0/\Delta}\exp(-2\Delta/\omega_0)$.

\bibliography{wwbib}

\end{document}